\documentclass{elsarticle}

\usepackage{amsmath,amssymb}
\DeclareMathOperator*{\SIGMA}{\Sigma}
\usepackage{algorithm}
\usepackage{algorithmic}
\usepackage{booktabs}
\usepackage{longtable}

\journal{Journal}
\begin{document}

\begin{frontmatter}



\title{Energy Efficient Ant Colony Algorithms for Data Aggregation in Wireless Sensor Networks}

\author[add1]{Chi Lin}
\author[add1]{Guowei Wu}
\ead{wgwdut@dlut.edu.cn}
\author[add1]{Feng Xia}
\ead{f.xia@acm.org}
\author[add1]{Mingchu Li}
\author[add1]{Lin Yao}
\author[add1]{Zhongyi Pei}
\address[add1]{School of Software, Dalian University of Technology, China}

\begin{abstract}
In energy-constrained wireless sensor networks, energy efficiency is critical for prolonging the network lifetime. A family of ant colony algorithms called DAACA for data aggregation are proposed in this paper. DAACA consists of three phases: initialization, packets transmissions and operations on pheromones. In the transmission phase, each node estimates the remaining energy and the amount of pheromones of neighbor nodes to compute the probabilities for dynamically selecting the next hop. After certain rounds of transmissions, the pheromones adjustments are performed, which take the advantages of both global and local merits for evaporating or depositing pheromones. Four different pheromones adjustment strategies which constitute DAACA family are designed to prolong the network lifetime. Experimental results indicate that, compared with other data aggregation algorithms, DAACA shows higher superiority on average degree of nodes, energy efficiency, prolonging the network lifetime, computation complexity and success ratio of one hop transmission. At last, the features of DAACA are analyzed.
\end{abstract}


\end{frontmatter}

\section{Introduction} \label{introduction}
The main task of wireless sensor network(WSN) \cite{label-19, label-20} is to detect and report the events of the physical world. In most cases, nodes are batteries powered \cite{label-21} with limited energy resources. Suppose that when a node runs out of power and stops working, the original transmission paths will be changed. Nodes nearby will suffer from heavier work, because of sharing responsibility of the exhausted node which casts heavy burden of them. The energy dissipating rate of these nodes will become faster. This process spreads which will cause the packets loss or even network congestion. Moreover, the performance of the network depends on the persistence of the sensors to a large extent. Hence, the main challenge for the energy-constrained network is to design energy-efficient routing protocols \cite{label-1, label-2, label-3, label-4, label-5} which guarantees the persistence and balances the energy consumption of the network.

In recent years, methods of data aggregation \cite{label-16} are attracting the attentions of the researchers. A large number of works of data aggregation protocols in wireless sensor networks for energy preservation have been published.

To begin with, we will provide a brief background of the data aggregation methods of WSN. The key idea is to combine the data coming from different sources which eliminates redundancy \cite{label-33}, minimizes the number of transmissions and thus saves energy. This method shifts the focus from the traditional address centric approaches for networking to a data centric approach.

Some existing methods \cite{label-6,label-7,label-8,label-10} focus on establishing and maintaining topology of the data aggregation tree in which data packets are transmitted and aggregated to specific nodes. In \cite{label-6}, the authors propose a method which generates Minimum Spanning Tree (MST) by implementing Prim algorithm from a global view. In \cite{label-7}, each node locally forms MST to establish the aggregation topology to eliminate energy costs. In \cite{label-8}, the authors integrate LMST with RNG\cite{label-9} to establish the network topology. In \cite{label-10}, the authors put forward an ant colony based data aggregation method which takes advantages of pheromones to heuristically set up the network topology. In the aforementioned methods, packets are transmitted in pre-allocated paths and aggregated in some specific nodes. Nevertheless, such topologies consume much energy in maintenance. Nodes in the aggregation tree are actually facing large energy consumption. The common problems of the existing methods are:

\begin{enumerate}
\item The energy consumptions of constructing and maintaining network topologies are high.
\item Issues of balancing energy consumption\cite{label-31, label-32} are not considered.
\end{enumerate}

The ant colony optimization(ACO)\cite{label-34} is one of the most useful swarm intelligence\cite{label-35} which has been
successfully applied in many optimization problems such as TSP\cite{label-36}, CVRP\cite{label-37} as well as routing in wireless sensor networks\cite{label-38,label-39,label-40}. In this paper, a measure of data aggregation based on ant colony optimization is proposed which overcomes shortcomings mentioned in \cite{label-16,label-13,label-14,label-15,label-17}. We expand the ant colony algorithms \cite{label-12} by proposing a family of energy-efficient Data Aggregation Ant Colony Algorithms (DAACA for short), which contains Basic-DAACA(i.e. the Basic algorithm of DAACA), ES-DAACA(i.e. Elitist Strategy based DAACA), MM-DAACA(i.e. Max-Min based DAACA) and ACS-DAACA(i.e. Ant Colony System based DAACA).

DAACA include three phases: 1) the initialization, 2) packets transmissions and 3) operations on pheromones. Initially, all the sensors are deployed randomly, then they set up their routing tables in a self-organization manner. We consider each node as an artificial ant which is raised in ant colony optimization for solving the global optimization problem. The packets transmitted between nodes are considered as tools for updating or adjusting the pheromones. Each node estimates the remaining energy and the amount of pheromones to compute the probabilities for dynamically selecting the next hop. After certain rounds of transmissions, the adjustments of pheromones are performed, which originally combines the advantages of global and local adjustments for evaporating or depositing pheromones. Four different pheromones adjustment strategies are designed to achieve the global optimal in prolonging network lifetime.

The contributions of the paper can be summarized as follows:
\begin{enumerate} \label{contribution}
\item We proposed a family of ant colony algorithms for data aggregation which aims at saving energy and prolonging network lifetime.
\item We devise the evaporating and the depositing pheromones approaches which take advantages of global and local merits.
\item We design three heuristic methods(i.e. ES-DAACA,MM-DAACA,ACS-DAACA) for improving the performance of the basic algorithm(i.e. Basic-DAACA) which specialize the way of evaporating and depositing pheromones.
\item We develop a platform to compare the characteristics(e.g. average energy cost, network lifetime, average degree and so on) of DAACA family with other data aggregation methods(e.g. PEDAP, PEDAP-PA, L-PEDAP and so on).
\end{enumerate}

The remainder of the paper proceeds as follows. Section \ref{relatedWork} gives an overview of the state of the art of data aggregation in wireless sensor networks. Section \ref{Preliminaries} introduces the preliminaries of the paper. Section \ref{SystemModel} presents the system model and defines the problem. Section \ref{AntColonyAlgorithmforDataAggregation} describes DAACA in details. The simulation results and performance analysis are presented in section \ref{simulationResultsandAnalysis}. Section \ref{conclusion} concludes the paper.

\section{Related Work} \label{relatedWork}
In literatures, many algorithms are proposed to prolong the lifetime of the network based on data aggregation \cite{label-29, label-30, label-31, label-6,label-7,label-8,label-10}. Some references are concentrating on establishing energy efficient topologies to save energy.

In \cite{label-6}, an algorithm named PEDAP (Power Efficient Data gathering and Aggregation Protocol) is proposed, which generates Minimum Spanning Tree (MST) by utilizing Prim algorithm. It is a near optimal minimum spanning tree based routing scheme. But it cannot guarantee the energy efficiency of the network, then an energy-aware version of PEDAP, namely PEDAP-PA is proposed to maximize the network lifetime. However, the sink node needs to periodically broadcast and calculate the MST of the network, thereby, the workload of the sink node is high.

In \cite{label-7}, a Local Minimum Spanning Tree algorithm called LMST is presented to establish the network topology, although it can effectively reduce the average degree of nodes, some prominent problems still emerge. Each node needs to periodically calculate and update the MST(Minimum Spanning Tree) locally which leads to a high computational overhead for each node. Moreover, each node needs to communicate with its neighbors to obtain the energy condition of neighbors, which still costs much energy.

In \cite{label-8}, a localized, self organizing, robust and energy efficient data aggregation algorithm named L-PEDAP is proposed which combines LMST with RNG \cite{label-9}. Although it is proved to have the capability of prolonging the network lifetime, its topology construction procedure is nearly identical with that of LMST, hence, it cannot be considered as an energy-efficient algorithm.

To reduce the cost of constructing and maintaining the energy-efficient topology, a heuristic method called ACA (Ant Colony Algorithm) which is based on ant colony algorithm is proposed in \cite{label-10}. In the ant colony optimization, a colony of artificial ants is used to construct solutions guided by the pheromones trails and heuristic information \cite{label-11}. This behavior enables ants to find the shortest paths between the food source and the nest in a random manner. The author initiatively designs the rules for depositing and evaporating pheromones. However, the requirements of depositing pheromones can be easily met. When depositing pheromones, each node needs to communicate with its neighbors. Therefore there are lots of communications within the pheromones adjusting period, which leads to much energy cost overhead. Therefore, it cannot be considered as an energy-efficient algorithm either.

Moreover, although some other methods do not depend on energy efficient topologies to save energy, some literatures are still valuable to be mentioned. In \cite{label-30}, the authors describe the design and implementation of a running system for energy-efficient surveillance which allows a group of cooperating sensor to detect and track the positions of moving vehicles. That method can trade off the energy-awareness and surveillance performance by adjusting the sensitivity of the system. In \cite{label-29}, the data aggregation is motivated by optimizations of aggregation queries. The authors present variety of techniques to improve the reliability and performance of the proposed solutions. In \cite{label-31}, the authors explore two methods to further reduce energy consumption in the context of network aggregation in sensor networks. Firstly, a group-aware network configuration method is designed, which groups the sensors into clusters. Secondly, a framework to use temporal coherency tolerances in conjunction with aggregation to save energy is proposed.

\section{Preliminaries} \label{Preliminaries}
In this section, we illustrate related knowledge in this paper.
\subsection{Ant Colony Optimization} \label{AntColonyOptimization}
The idea of Ant Colony Optimization is originated by Marco Dorigo\cite{label-41}, in observing the exploitation of seeking shortest path between their nest to the food source although limited cognitive abilities are owned by ants. In Ant Colony Optimization, artificial ants are implemented to search for heuristic solutions of the problems. Originally, artificial ants are utilized to find the solutions stochastically based on the probabilities whose calculations are defined according to certain circumstances(i.e. according to the characteristic of the problem). After a feasible solution is constructed, the evaporation and releasing of pheromones are executed which dynamically update the amount of pheromones. Therefore, the solution will gradually approach to the global optimal solution of the problem. This procedure repeats until certain number of ants have finished their trails. The process of ACO can be summarized as Algorithm \ref{alg:ACOPseudocode}.
\begin{algorithm}
\caption{ACO Pseudocode} \label{alg:ACOPseudocode}
\begin{algorithmic}[1]
\STATE Initialization of data structure and parameters
\WHILE {terminate conditions}
\STATE Construct solutions
\STATE Update statistics
\STATE Update pheromones
\ENDWHILE
\end{algorithmic}
\end{algorithm}

In general, we can summarize the advantages of ant colony optimization.
\begin{enumerate}
\item  ACO is a probabilistic mechanism for solving computational problems which can be reduced to finding good paths in graphs.\label{item1}
\item The solution of ACO will gradually approach to the global optimal solution in terms of adjusting the quantities of pheromones.\label{item2}
\item In ACO, each ant is considered to have individually limited cognitive ability.
\item The process of ACO is the cooperation of ants which gradually find optimal solutions in graph.
\end{enumerate}

In wireless sensor networks, a large number of nodes are deployed. To find the global optimal prolong lifetime solution can be transformed into forming data aggregation tree by global optimization. Hence, firstly, our aim is to find a solution which can approach to global optimization. Secondly, to prolong the lifetime of the network, we need to balance the energy cost of each node, therefore, we need to establish dynamical topology of the networks. Thirdly, each node has limited computation capability, it only hold the local information(i.e. the information of neighbors). What's more, the network can be abstracted as a graph where the vertexes stand for the nodes and the edges represent the link between nodes. Last but not least, researches of the ACO in wireless sensor networks\cite{label-10,label-42,label-43} lay a good foundation for our work. From the analysis above, we conclude that ACO is suitable for data aggregation in wireless sensor network. Therefore, we employ ant colony optimization in resolving data aggregation in wireless sensor network.

\subsection{Ant Colony System} \label{AntColonySystem}
Ant Colony System(ACS) applies the mechanisms of ACO but some changes has been made to overcome the drawbacks of ACO and enhance the performance of ACO.
The main differences can be summarized as:
\begin{enumerate}
\item It applies a more active behavior, which can make full use of the accumulated pheromones in the path.
\item The evaporating and releasing of the pheromones are only carried out in the most optimal path so far.
\item When an ant passes through a trail, the pheromone of that trail will be reduced which aims at enhancing the possibility of finding more optimal solutions in other trails.
\end{enumerate}

Inspired by the features of ACS, we implement the ant colony system into our data aggregation method in section \ref{ACSDAACA} to enhance the performance.

\subsection{ACA Algorithm} \label{ACAAlgorithm}
In \cite{label-10}, the authors propose an data aggregation(ACA). First, each compute the hop count to the sink node. When the source node wants to send data, it will select the next hop node by the probability which mainly depends on hop count of the neighbor nodes to the sink.
\begin{equation}
p_{k}(i,j)=\frac{\tau(i,j) \times \eta(i,j)^{\beta}}{\Sigma_{u \in N_i}\tau(i,u) \times \eta(i,u)^{\beta}} \nonumber
\end{equation}
where $\tau(i,j)$ is the pheromones level from node $i$ to node $j$, $\eta(i,j)$ is the inverse of the hop count from node $j$ to the sink node adding one, $N_i$ is the number of neighbors of the node $i$, and $\beta$ is a parameter which determines the relative influence of heuristic values $\eta(i,j)$, $\beta > 0$.
Then the authors design the message communication mechanism which increase the probability of intersection of the routing path to promote the results of data aggregation.
Finally, the evaporation and releasing of pheronmones are defined which dynamically adjust the quantities of pheromones in the paths. A node who has a lower hop count and happen to exist in the path are more likely to get pheromone which enhances the probability of obtaining messages in later transmissions. Whereas a node cannot receive a message for more and certain round, the pheromones of such node will be evaporate, which reduce the possibility of receiving of nodes of such nodes.

To aggregate size of packets, in this paper, we develop a family of ant colony based data aggregation algorithm DAACA. The similarities and differences between DAACA family and ACO can be summarized as follows:
\textbf{Similarities:}
\begin{enumerate}
\item Ant Colony Optimization methods are employed which aims at establishing data aggregation tree.
\item Methods of releasing and evaporating of pheromones are developed.
\item The ultimate targets are to save energy.
\end{enumerate}
\textbf{Differences:}
\begin{enumerate}
\item DAACA focus on balancing the energy cost of the whole network to prolong the network lifetime whereas ACO does not take the lifetime into account.
\item When selecting the next hop, the remaining energy, the distance between the neighbor to the sender are considered which aims at fairly selecting the next hop based on energy cost in DAACA, whereas ACA only cares about the distance and the quantities of the pheromones.
\item In ES-DAACA, MM-DAACA and ACS-DAACA, we take advantages of accumulated pheromones to increase the probability of selecting the global optimal solution. But ACO does not take any actions to obtain global optimal solutions.
\item In MM-DAACA and ACS-DAACA, the range of the quantities of pheromones are constrained which aims at avoiding trapping into local optimal solutions whereas ACO does not limit the ranges of pheromones.
\item In ACS-DAACA, we reduce the probabilities of the nodes who locates in the aggregation tree to diversify the solutions to get more optimal solutions(i.e. the data aggregation tree). On the contrary, ACO can not guarantee the diversity of the solutions.
\end{enumerate}

\subsection{Energy Consumption Model} \label{energyConsumptionModel}
Many energy models \cite{label-15} are used for energy consumption in wireless sensor networks. In our work, we employ the model in which a packet with the size $k$ to be transmitted in the distance of $d$ will consume:

\begin{equation}
E_{Tx}(k,d)=E_{Tx-elec} \times k + \varepsilon_{amp} \times k \times d^2
\end{equation}
for the transmitter and
\begin{equation}
E_{Rx}(k)=E_{Rx-elec} \times k
\end{equation}
for the receiver respectively. The definitions of the symbols are listed in table \ref{tabel1}.

\begin{table}
\centering
\caption{Energy Dissipated Table} \label{tabel1}
\begin{tabular}{c|c|c}
\hline
\textbf{Symbol} & \textbf{Meaning} & \textbf{Energy Dissipated} \\
\hline
$E_{Tx-elec}$ & Transmitter Electronics & 50nJ/bit \\
$E_{Rx-elec}$ & Receiver Electronics & 50nJ/bit \\
$\varepsilon_{amp}$ & Transmit Amplifier & 100J/bit/{$m^2$} \\
\hline
\end{tabular}
\end{table}

\section{System Model} \label{SystemModel}
We consider a scenario where the sensor nodes are homogenous and energy-constrained. All the nodes (including the sink node) are stationary and randomly distributed. Further, replenishing energy via replacing batteries on hundreds of nodes (in possibly harsh terrains) is infeasible. The basic operation in such a system is the systematic gathering of sensed data to be eventually transmitted to a base station for processing. The key challenge in such data gathering is conserving the sensor energies, so as to maximize their lifetimes.

\subsection{Definitions and Assumptions}
In this section, some definitions and assumptions are defined below.
\subsubsection{Definition}
\textbf{Remaining Energy}: The initial energy minus the energy consumed in transmission process.

The remaining energy reflects the usage of energy. Moreover it can indicate when the network exhausts.

\textbf{Energy Difference}: The maximum remaining energy minus the minimum remaining energy in the network.

The energy difference indicates the balanced usage of energy in the network. Small energy difference shows the balanced usage of energy of all nodes in the network and vice versa.

\subsubsection{Assumption}

\textbf{Assumption 1}: In one aggregation procedure, $n$ packets with the size of $k$ will be eventually compressed into one with the size of $k$.

If the packet size is bigger than $k$, it will be sliced into pieces with constant size of $k$. If the size if smaller than $k$, it will be enlarged into size of $k$. This assumption is identical with that of \cite{label-11}

\textbf{Assumption 2}: Only the transmission process costs energy.

Since we mainly focus on the energy cost in data aggregation to prolong the lifetime, therefore we omit effects of the energy costs of nodes of their own(e.g. calculating, waiting and so on). We assume that the energy cost can only happen when transmitting or receiving packets.

\subsection{Problem Statement} \label{ProblemStatement}
The network is modeled as a visibility graph $G=(V, E)$, where $V$ is the set of sensor nodes. Each node has the maximum transmission range denoted by $R$, by which it can setup its neighbor set and routing table. $e_{ij}$ represents the distance between node $i$ and node $j$ which is smaller than $R$, meanwhile the neighbor set (denoted as $Nb(i)$ ) of node $i$ indicates a set of nodes whose distances are less than $R$(i.e. $Nb(i)=\{j|e_{ij} \leq R\}$). The set of $e_{ij}$ comprises the $E$. Each source periodically senses and collects the data from the surroundings, then sends the data to the next hop until the data reaches the sink node. The goal is to find an energy efficient routing path to prolong the network lifetime. The best way to solve this problem is to aggregate the multiple data packets and reduce the traffic in the network. Thus, multiple transmission paths should be merged into smaller ones to relieve the energy consumptions of data transmissions. Multiple sources periodically send data to their neighbors until the data packets reach the sink node in a hop by hop manner.

\section{Ant Colony Algorithms for Data Aggregation} \label{AntColonyAlgorithmforDataAggregation}

Overall we propose the system architecture of DAACA. Then we design a basic algorithm, after that three heuristic algorithms are proposed to enhance the performance of the basic algorithm.

Since each algorithm in DAACA family has the same algorithm structure, differences only exist in the process of evaporating and depositing pheromones. We firstly illustrate algorithm structure of DAACA family and then introduce each algorithm in details. The symbols used in DAACA are illustrated in table \ref{tab:definitions}.

\begin{center}
\begin{longtable}{|c|p{6cm}|}
\caption{Related Definitions} \label{tab:definitions} \\\hline
\multicolumn{1}{|c|}{\textbf{Symbol}} & \multicolumn{1}{c|}{\textbf{Notations}} \\
\hline \endfirsthead \multicolumn{2}{c}{{\bfseries \tablename\ \thetable{} -- continued from previous page}} \\
\hline \multicolumn{1}{|c|}{\textbf{Symbol}} &\multicolumn{1}{c|}{\textbf{Notations}} \\
\hline \endhead \hline \multicolumn{2}{|r|}{{Continued on next page}} \\
\hline\endfoot \hline \hline\endlastfoot
$R$ & The maximum transmission range. \\
$id$ & The identity of the node. \\
$Nb(i)$ & The neighbor nodes of node $i$. \\
$round$ & The transmission times. \\
$roundToUpdate$ & The number of rounds to update the pheromones. \\
$p(i,j)$ & The probability for node $i$ to select node $j$ as the next hop.\\
$\eta(i,j)$ & The pheromone of node $i$ to node $j$.\\
$E'_{distance}(i,j)$ & The energy distance from node $i$ to node $j$.\\
$E_{distance}(i,j)$ & The Euclidian distance between node $i$ and node $j$. \\
$e_1(i)$ & The current value of node $i$ divided by $E_{init}$. \\
$e_2(i,j)$ & The estimation energy of node $i$ to node $j$ divided by  $E_{init}$. \\
$e_{ij}$ & The distance between node $i$ and node $j$. \\
$E_{cur}(i)$ & The current energy of node $i$.\\
$E_{estimate}(i,j)$ & The evaluation energy of node $i$ to node $j$.\\
$E_{init}$ & The initial energy of each node. \\
$\tau(i,j)$ & The inverse value of $E'_{distance}(i,j)$ \\
$Times(i,j)$ & The transmission times from node $i$ to node $j$. \\

$M_{PCost}$ & The minimum energy cost in a path. \\
$\eta_{Max}$ & Upper bound of pheromones. \\
$\eta_{Min}$ & Lower bound of pheromones. \\
$conj(i)$ & The conjunction times of node $i$. \\
$source$ & The source node($s$ for short) of the network. \\
$destination$ & The destination node ($d$ for short) of the network. \\
$current$ & The current node($c$ for short) who receives the data packet. \\
$next$ & The next node($n$ for short)which current will send the packet to.\\
$P_{min}$ & The minimum energy cost path. \\
\end{longtable}
\end{center}

\subsection{DAACA Algorithm Structure} \label{DAACAAlgorithmStructure}

Similarly, in wireless sensor network, nodes are considered as the artificial ants. The amount of the pheromones of links are recorded in the routing table, whenever a node receives a packet, it will figure out the next hop based on the information of the pheromones, the distance and the remaining energy according to the routing table. Then each node will update the information. After certain rounds of transmissions, procedure of evaporating and depositing pheromones will be applied which help DAACA to adjust the amount of the pheromones. The objective of this process is to guide the aggregation routing path to approach to the global optimal routing path which will conserve energy for the network to some extent. Next we will illustrate the realizations of DAACA in details.

Initially every node constructs its own neighbor set and routing table by broadcasting its $id$ (i.e. the identity of the node) and location information within $R$. Afterwards, the transmission process starts. Packets are sent from the source nodes to the sink node in each round. When a node receives a packet, it will evaluate contents of the routing table (e.g. the remaining energies, the amount of pheromones) to calculate the transmission probabilities of selecting the next hop. Then one of its neighbors will be selected. When the number of $round$ equals to the $roundToUpdate$, all nodes will perform the evaporating and depositing operations. Each node evaporates the pheromones of its neighbor and deposits pheromones according to different specific conditions (referring to section \ref{BasicDAACA}- \ref{ACSDAACA} in the following sections). After that each node will update the pheromones of its neighbors.

\begin{algorithm}
\caption{DAACA Structure} \label{DAACAStructure}
\begin{algorithmic}[1]
\STATE Network Initialization \\ \label{networkInitialization}
Node Initialization \\
Neighbor Initialization \\
Routing Table Initialization \\
\STATE The source node begin to send its data packets to the destination hop-by-hop.
\FORALL {$t \in Nbr(s)$} \label{tranmissionStart}
\STATE $s$ evaluates the energy of $t$.
\STATE $s$ calculates $p(s,t)$.
\STATE $s$ selects the next hop node $n$ based on $p(s,t)$
\ENDFOR
\STATE $s\longrightarrow n$ $//$ $s$ sends the packets to $n$
\WHILE {$s \neq d$}
\STATE $c = n$
\STATE $c$ aggregates data packets
\FORALL {$t \in Nbr(c)$}
\STATE $c$ evaluates the energy of $t$.
\STATE $c$ calculates $p(c,t)$.
\STATE $c$ selects the next hop node $n$ based on $p(c,t)$
\ENDFOR
\STATE $c \longrightarrow n$
\ENDWHILE \label{tranmissionEnd}
\STATE $round = round + 1$
\IF {$round$ = $roundToUpdate$} \label{pheromonesAdjustStart}
\STATE Evaporating Pheromones.
\STATE Depositing Pheromones.\label{pheromonesAdjustEnd}
\STATE Updating Routing Table. \label{updateRoutingTableStart}
\STATE Energy Broadcasting. \label{updateRoutingTableEnd}
\ENDIF
\end{algorithmic}
\end{algorithm}

Algorithm \ref{DAACAStructure} illustrates the structure of DAACA. Firstly, the initialization of the network is carried out. Each node broadcasts "hello" message to its neighbors. Then only the nodes who are nearer to the sink node will be recorded in the routing table and the contents of the table are initialized. Then the transmission begins, the source nodes periodically send the data packets to the sink node. When a relay node receives a packet, it will calculate the probability of selecting the next hop based on the elements in the routing table (e.g the distance, the amount of the pheromones, estimated remaining energy et.al). The transmission proceeds until the data are forwarded to the sink node (i.e. line \ref{tranmissionStart}-line \ref{tranmissionEnd}). After transmitting $roundToUpdate$ rounds, adjustments of pheromones are carried out (i.e. line \ref{pheromonesAdjustStart}-line \ref{pheromonesAdjustEnd}). Each node updates the routing table to keep the information fresh. At last each node broadcasts its actual energy, by which the neighbor nodes will update of the contents of the routing table (i.e. line \ref{updateRoutingTableStart}-line \ref{updateRoutingTableEnd} ).

\begin{figure}
\centering
  \includegraphics[width=10cm]{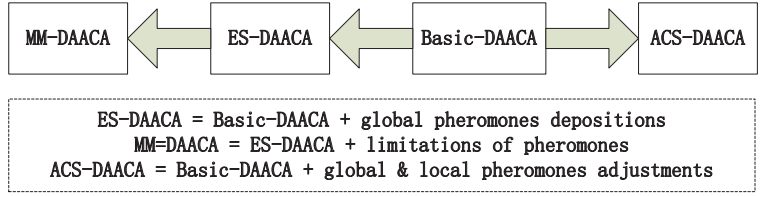}\\
  \caption{The relationship between Basic-DAACA, ES-DAACA, MM-DAACA, ACS-DAACA}\label{fig:relation}
\end{figure}

In the following sections, we present four ant colony heuristic algorithms (i.e. Basic-DAACA, ES-DAACA, MM-DAACA, ACS-DAACA) for adjusting the pheromones especially the evaporating and depositing of pheromones. Basic-DAACA is the basic algorithm, other algorithms are designed based on Basic-DAACA. The relationship between Basic-DAACA, ES-DAACA, MM-DAACA and ACS-DAACA can be formalized in figure \ref{fig:relation}.

ES-DAACA adds additional global pheromones depositions into Basic-DAACA. MM-DAACA limits the ranges of pheromones in ES-DAACA. ACS-DAACA adds both global and location pheromones adjustments in Basic-DAACA. Followings are explicit illustrations of DAACA family.

\subsection{Basic-DAACA} \label{BasicDAACA}
Basic-DAACA is the basic algorithm of DAACA family. For the pheromone is the most critical part in adjusting the probabilities in the routing table, we firstly illustrate the constitution of the routing table. The routing table contains the following elements: \{ $id$, $E'_{distance}(i,j)$, $E_{estimate}(i,j)$, $\eta(i,j)$, $p(i,j)$ \}, $j\in Nb(i)$. $id$ refers to the identity of the node. $p(i,j)$ is the probability for node $i$ to select node $j$ as the next hop which is computed as follows:

\begin{equation} \label{e1}
p(i,j)= \frac{{\tau(i,j)}^{\alpha} \times {\eta(i,j)}^{\beta}}{
\SIGMA \limits_{j \in Nbr(i)}
[{\tau(i,j)}^{\alpha} \times {\eta(i,j)}^{\beta}]}
\end{equation}

$\alpha$ and $\beta$ are two parameters which determine the relative influence of $\tau(i,j)$ and $eta(i,j)$. $\tau(i,j)$ represents the inverse value of the energy distance $E'_{distance}(i,j)$ between node $i$ and $j$, which holds:

\begin{equation}
\tau(i,j)= \frac{1}{E'_{distance}(i,j)}
\end{equation}

$E'_{distance}(i,j)$ can be computed as:
\begin{equation}
E'_{distance}(i,j)= \frac{E_{distance}(i,j)}{e_{1}(i) \times e_{2}(i,j)} (0<e_1<1, 0<e_2<1)
\end{equation}

Where:
\begin{equation}
e_1(i)=\frac{E_{cur}(i)}{E_{init}}
\end{equation}

\begin{equation}
e_2(i,j)=\frac{E_{estimate}(i,j)}{E_{init}}
\end{equation}

$E_{distance}(i,j)$ is the energy distance which can be calculated as:
\begin{equation}
E_{distance}(i,j)=E_{Tx-elec} \times k + \varepsilon_{amp} \times k \times {e_{ij}}^2
\end{equation}

$k$ is the size of the packet. $E_{estimate}(i,j)$  denotes the energy of node $j$ estimated by node $i$, which can be estimated according to:

\begin{equation}
E_{estimate}(i,j)= E_{init} - \frac{E_{init} - E_{estimate}(i,j)}{Times(i,j)} \times [Times(i,j) + 1]
\end{equation}

When a node receives a packet, it will evaluate the remaining energy of all the neighbors and updates all the values in the routing table to dynamically select the next hop.

When $round$ is multiple of $roundToUpdate$, procedure of adjusting the pheromones starts. First, the pheromones should be evaporated. We utilize the equation (\ref{e7}) to evaporate pheromones.

\begin{equation} \label{e7}
\eta(i,j)= (1- \rho) \times \eta(i,j)
\end{equation}

$\rho$ stands for the fraction of pheromones that are not evaporated. Then the procedure of depositing pheromones is performed. Each node selects the neighbor with the maximum estimation energy (e.g. the node $j$), and increases the pheromone of node $j$ by $E_{distance}(i,j)$ as:
\begin{equation} \label{e8}
\eta(i,j)= \eta(i,j) + E_{distance}(i,j)
\end{equation}

When $round$ is multiple of $roundToUpdate$, algorithm \ref{alg2} is called to adjust the pheromones for each node. Firstly, the evaporating and depositing of pheromones are taken(i.e. line \ref{alg2EvaporatingDepositingStarts} - line \ref{alg2EvaporatingDepositingEnds}). Then updating of the routing table are carried out(i.e. line \ref{alg2UpdateRoutingTableStarts} - line \ref{alg2UpdateRoutingTableEnds}).

Once algorithm \ref{alg2} is finished, in the next period of $roundToUpdate$, the node with the highest $E_{distance}(i,j)$ will have a higher probability of being selected as the next hop. After finishing depositing pheromones, the process of updating the estimated energy value is performed. Each node will broadcast its current energy in the range of $R$. The value of $E_{estimate}(i,j)$ will be updated.

If a node happens to exist in the conjunction of two or more different paths, the parent nodes will also use equation (\ref{e8}) to increase the pheromones.

\begin{algorithm}
\caption{Adjusting pheromones of Basic-DAACA} \label{alg2}
\begin{algorithmic}[1]
\FORALL {$i \in V$} \label{alg2EvaporatingDepositingStarts}
\STATE $i$ evaporates the pheromones according to equation (\ref{e7}).
\STATE $i$ searches for the node with the highest $E_{estimate}(i,j)$ where  $j \in Nb(i)$ in neighbor set.
\STATE $i$ deposits pheromones according to equation (\ref{e8}).
\IF {$conj(i) \geqslant 2$}
\FORALL{$t \in Nb(i)$}
\STATE $t$ deposits pheromones of $i$ according to equation (\ref{e8}).
\STATE $i$ broadcasts the current energy.
\ENDFOR \label{alg2EvaporatingDepositingEnds}
\ENDIF
\IF {$i$ receives a broadcast message from $j$} \label{alg2UpdateRoutingTableStarts}
\STATE $i$ updates $E_{estimate}(i,j)$.
\STATE $i$ updates the routing table.
\ENDIF \label{alg2UpdateRoutingTableEnds}
\ENDFOR
\end{algorithmic}
\end{algorithm}
Basic-DAACA algorithm avoids the computing overhead of repeatedly constructing or maintaining the network topology, which saves energy and prolongs the network lifetime. In the following sections, some heuristic methods are proposed to enhance the performance of Basic-DAACA.

\subsection{Elitist Strategy based DAACA(ES-DAACA)} \label{ESDAACA}
In order to establish better aggregation paths, the global pheromones deposition is employed. We call this heuristic method Elitist Strategy based DAACA(ES-DAACA for short). We apply another format of the packet header, which obtains the $id$ sequence of nodes and total energy consumption as: {$ID\_List$, $Energy\_Consumption$}. $ID\_List$ records all the nodes in the trace from the source to the current node. $Energy\_Consumption$ is defined as the total transmission energy costed from the source to the current node.

Once a packet is received by a node, algorithm \ref{alg3} will be taken which adds $id$ information and energy consumption information into the newly defined packet header. Owing to the header, the sink node can find the minimum cost aggregation path and adjust the amount of pheromones to obtain more optimal solutions.
\begin{algorithm}
\caption{Packet receiving process of ES-DAACA} \label{alg3}
\begin{algorithmic}[1]
\IF {$i$ receives the data packet} \label{alg3Starts}
\STATE $i$ selects the next hop $j$.
\STATE $i$ adds its id information into $ID\_List$.
\STATE $i$ adds $E_{distance}(i,j)$ into the energy consumption field. \label{alg3Ends}
\ENDIF
\end{algorithmic}
\end{algorithm}

When a packet is sent to the sink node. The sink node checks if the $Energy\_Cosumption$ is smaller than $M_{PCost}$, if so, the $M_{PCost}$ and the corresponding $ID\_List$ will be updated. When $round$ is multiple of $roundToUpdate$, the sink node will perform global deposition process. Each node in the $ID\_List$ will add additional $M_{PCost}$ pheromones as in equation (\ref{e9}). Therefore, ES-DAACA can increase the amount of phermones of those nodes who locate in the path with the minimum energy consumption. More optimal solutions of data aggregation paths will be likely to be built. Obviously ES-DAACA will enhance the performance of Basic-DAACA.

\begin{equation}
\eta(i,j)= \eta(i,j) + M_{PCost} (j \in Nb(i), i \in P_{min}, j \in P_{min}) \label{e9}
\end{equation}

Therefore, ES-DAACA will enhance the probability of finding global optimization solution, which preserves the energy and prolongs the network lifetime.

\subsection{Maximum \& Minimum based DAACA (MM-DAACA)} \label{MMDAACA}
To limit the bound of the pheromones, we impose two constants to represent the maximum and minimum bounds of pheromones. We name this heuristic algorithm as Maximum \& Minimum based DAACA (MM-DAACA for short).

In MM-DAACA, all the steps are identical with ES-DAACA, the only difference is that we use [$\eta_{Min}$, $\eta_{Max}$] to limit the range of the pheromones. The reason is that if the pheromone is not limited in a range, some paths will own higher probabilities than the others, nevertheless, with the transmission going on, nodes in this kind of path will cost more energy than the others. But according to equation (\ref{e1}), the amount of pheromones is still large, which may cause them more likely to be selected as the next hop, and ultimately results in local optimal solution. However, if the constraints of pheromones are imposed, the aforementioned phenomenon can be avoided, variety of paths will be formed, hence, the global optimal solution will be more likely to be found.

\subsection{Ant Colony System based DAACA(ACS-DAACA)} \label{ACSDAACA}
Similar to the ant colony system mentioned in section \ref{AntColonySystem}, we develop the Ant Colony System based DAACA (ACS-DAACA for short) algorithm for establishing the data aggregation paths to preserve energy.

In ACS-DAACA, when selecting the next hop node, the algorithm \ref{alg4} is performed.
\begin{algorithm}
\caption{Selecting the next hop of ACS-DAACA} \label{alg4}
\begin{algorithmic}[1]
\STATE Generate a random number $q$ in [0, 1].
\IF {$q \leqslant q_0$ }
\STATE Select the neighbor node with the maximum ${\tau(i,j)}^{\alpha} \times {\eta(i,j)}^{\beta}$
\ELSE
\STATE {select the next hop according to (1)}
\ENDIF
\end{algorithmic}
\end{algorithm}
$q_0$ is a parameter. After successfully sending the packet in each round, each node will locally update the pheromones as follows.
\begin{equation} \label{e10}
\eta(i,j)= (1- \rho) \times \eta(i,j) + \rho \times M_{PCost} (j \in Nb(i), i \in P_{min}, j \in P_{min})
\end{equation}
Since the result of equation (\ref{e10}) is proved to be constrained within a certain range \cite{label-31}, therefore, it is similar to MM-DAACA which can maintain the variety of the solutions. What's more, in ACS-DAACA, we propose local pheromone updating strategy as :
\begin{equation} \label{e11}
\eta(i,j)= (1- \zeta) \times \eta(i,j) + \zeta \times {\Delta\eta(i,j)}^{best}
\end{equation}
$\zeta$ stands for the rate of evaporating pheromones.
Where
\begin{equation}
{\Delta\eta(i,j)}^{best} = \min \limits_{j \in Nbr(i)}({\eta(i,j)})
\end{equation}

The local pheromones updating aims at reducing the probability of the selected node, thus the probabilities of unselected nodes will increase. Consequently, ACS-DAACA will generate more aggregation paths than MM-DAACA through which global optimal solution will be more likely to be found.

\section{Simulation Results and Analysis} \label{simulationResultsandAnalysis}
In this section, several simulations are performed to evaluate and compare the performances of the DAACA family with other data aggregation algorithms. We use Java to develop a simulation platform to analyze the energy cost of each node and the network lifetime. We have testified the accuracy of the platform by simulating the third set of experiments in \cite{label-10} and get identical statistics on the platform. Then we simulate different algorithms(i.e. LMST, PEDAP, PEDAP-PA, L-PEDAP, ACA, Basic-DAACA, ES-DAACA, MM-DAACA, ACS-DAACA) on the platform. In the experiments, multiple sources periodically send data to the sink node. Related parameters are shown in table \ref{table3} which are identical with those in \cite{label-10}. We have run millions of experiments to highlight the advantages of DAACA in this paper. What's more, features of network topology are discussed. Furthermore, the time complexity analysis and characteristic analysis are given.

\begin{table}
\centering
\caption{Experimental Parameters} \label{table3}
\begin{tabular}{c|c}
\hline
\textbf{Item} & \textbf{Parameters} \\
\hline
Algorithm & \multicolumn{1}{p{6cm}}{LMST, PEDAP, PEDAP-PA, L-PEDAP, ACA, Basic-DAACA, ES-DAACA, MM-DAACA, ACS-DAACA} \\
Packet Size & 4098 bits \\
(Length$\times$Width, Node Number) & \multicolumn{1}{p{6cm}}{(40 $\times$ 50, 200), (60 $\times$ 60, 400), (70 $\times$ 80, 600), (80 $\times$ 90, 800), (100 $\times$ 100, 1000), (100 $\times$ 120, 1200), (100 $\times$ 140, 1400), (120 $\times$ 130, 1600), (130 $\times$ 120, 1800), (140 $\times$ 140, 2000)} \\
Node Distribution & Random \\
Initial Node Energy & 10J \\
Packets Number & \multicolumn{1}{p{6cm}}{1000, 2000, 3000, 4000, 5000, 6000, 7000, 8000, 9000, 100000} \\
Number of Sources & 10 \\
$R$ & 10 \\
$\alpha, \beta$ & $\alpha=2, \beta=2$ \\
Initial Pheromone & 0.8 \\
$\eta_{Min}$, $\eta_{Min}$ & $\eta_{Min}=0.5, \eta_{Max}=0.9$ \\
$\rho$ & $\rho=0.2$ \\
$roundToUpdate$ & 100, 200 \\
$\zeta$ & 0.9 \\
$q_0$ & 0.5 \\
$rho$ in ACA & 0.3\\
\hline
\end{tabular}

\end{table}

\subsection{Average Energy Cost} \label{averageEnergyCost}
\begin{figure}
\centering
  \includegraphics[width=10cm]{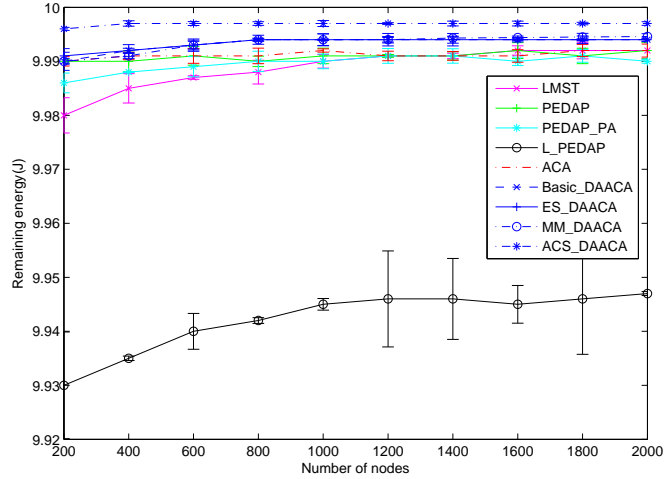}
  \caption{Average Energy Cost after Transmitting 10000 Packets} \label{figure1}
\end{figure}

All the nodes are deployed randomly as (Length $\times$ Width, Node Number). Different numbers of packets are transmitted from the source nodes to the sink node. We calculate the average remaining energy of nodes with different algorithms. Higher remaining energy means less energy cost. Figure \ref{figure1} - figure \ref{figure3} illustrate the comparison results. Firstly each algorithm will initialize the topology of the network. With the transmissions going on, some of the algorithms will re-construct (or maintain) the topology to balance the energy cost of each node \cite{label-32}. Hence, the topologies of some algorithms are dynamical.

\begin{figure}
\centering
  \includegraphics[width=10cm]{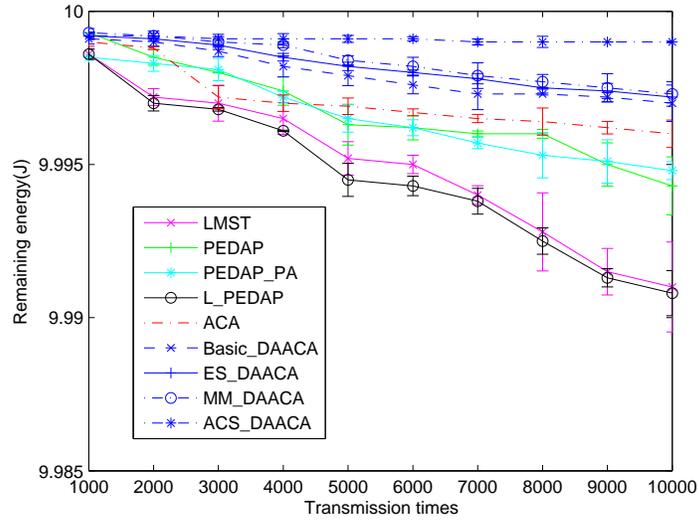}
  \caption{Average Energy Cost of the Network with the size (140 $\times$ 140, 2000)} \label{figure2}
\end{figure}

\begin{figure}
\centering
  \includegraphics[width=10cm]{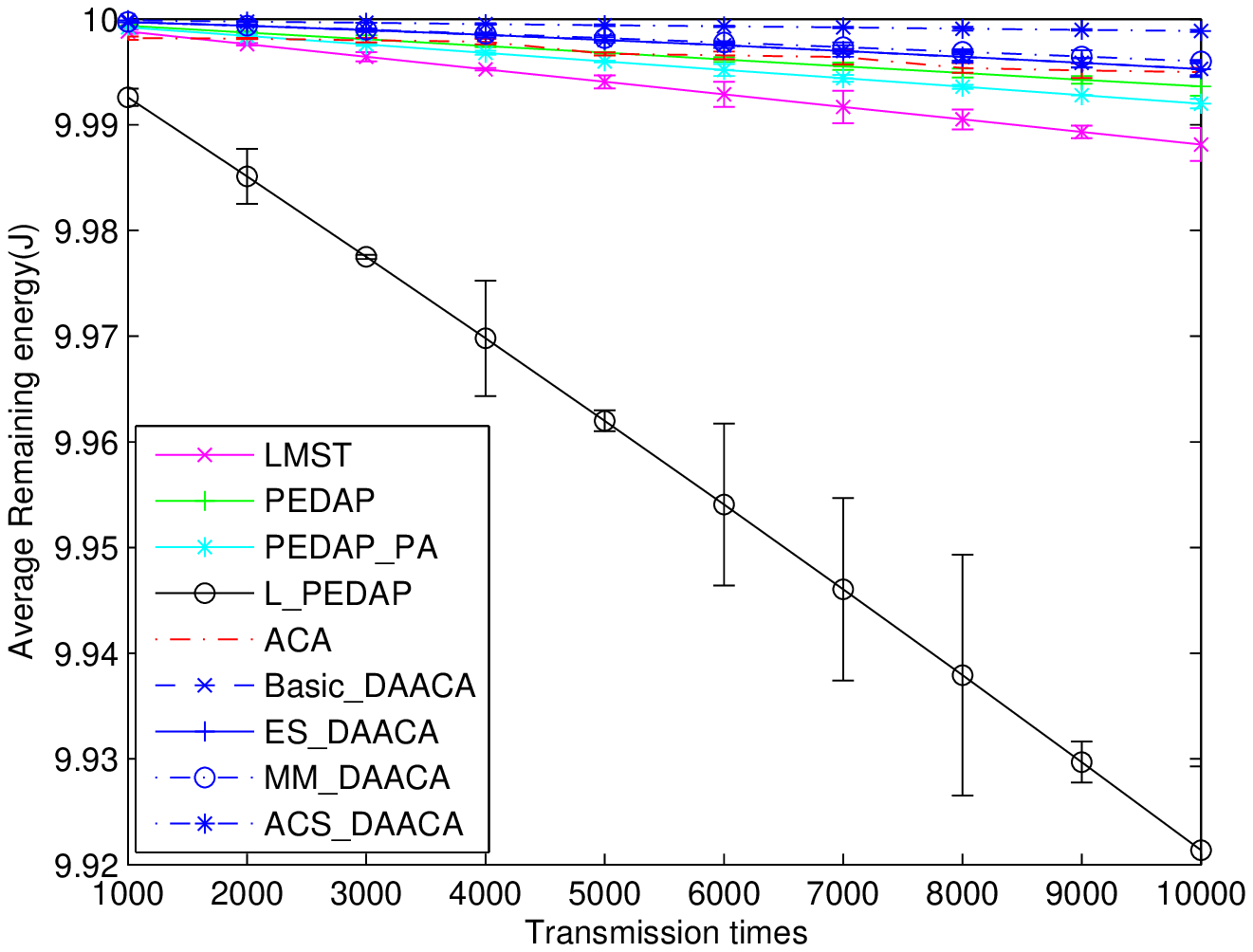}
  \caption{Average Results of all the Experiments of Average Energy Cost} \label{figure3}
\end{figure}

The mechanism of establishing network topology of LMST and L-PEDAP are nearly identical, the energy consumptions of construction and maintenance of the network topology are high. Because each node in the network establishes the minimum spanning tree in the neighborhood locally. Then local minimum spanning trees connect into the topology of the network. When establishing the local minimum spanning tree, the weight of edge is directly related to the remaining energy of the node.  Hence, each node needs to communicate with neighbors to obtain their remaining energies periodically to update the weights of edges locally, frequent communications cast a heavy burden of all the nodes in the network. Therefore, the LMST and L-PEDAP cost more energy than others.

In ACA, the energy cost concentrates on the frequent depositions of pheromones. Because whenever a packet is received by a node, if the $TTL$ of the packet is bigger than the $EHC$(Extended Hop Count)\cite{label-10} of that node, the deposition of pheromones will start. The feedback message will be transmitted to inform the upstream nodes to deposit pheromones. Suppose there are a large number of packets transmitting in the network, and $TTL>>EHC$, the feedback messages will be massively used, which consumes considerable energy. Therefore the energy consumption mainly concentrates on too frequent transmitting feedback messages.

In PEDAP and PEDAP-PA, the Prim algorithm are used to construct the topology and maintain the topology from the global view. Such methods can save the energy of communicating with neighbor nodes, but the sink node will suffer from more energy consumptions than the other nodes, because this is a centralized version of method. The sink node is responsible for establishing the topology of the network and informing all the nodes about the topology of the network via broadcasting. The energy cost of broadcast is high, especially when some nodes locate far away from the sink node. Whenever the network topology is changed, the broadcasting will be executed. Therefore, the energy costs of PEDAP and PEDAP-PA in constructing and maintaining the network topology are high.

In the family of DAACA, there is no energy consumption in maintaining the topology or reconstructing the MST, therefore, the average energy cost is lower than other algorithms. The energy consumption concentrates on sending packets to the next hop. After certain rounds of transmissions, the global pheromones adjustment will be carried out, which does not cost much energy. From the figure \ref{figure1} - figure \ref{figure3} we can see, the relationship among the four methods of the DAACA family in terms of average energy cost is ACS$ < $MM$ < $ES$ < $Basic. MM-DAACA, ES-DAACA and ACS-DAACA enhance the quality of the solutions of Basic-DAACA which generate more energy preservation data aggregation paths for the network and eventually save the energy.

\subsection{Energy Difference} \label{EnergyDifference}
In this section, we calculate energy difference to evaluate the balanced consumption of energy. We aims at testifying which method can balance the usage of energy. Because the balanced consumption can directly affect the lifetime of the network, which is the most prominent feature in energy efficiency in wireless sensor networks. If the energy difference is big, the energy is not averagely used whereas smaller energy difference indicates the balanced energy consumption.
\begin{figure}
  \centering
  \includegraphics[width=10cm]{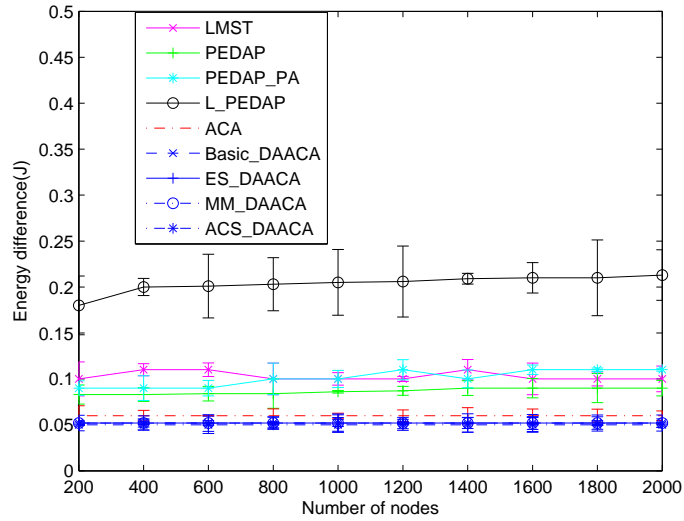}\\
  \caption{Energy Difference after Transmitting 10000 Packets}\label{figure4}
\end{figure}

\begin{figure}
  \centering
  \includegraphics[width=10cm]{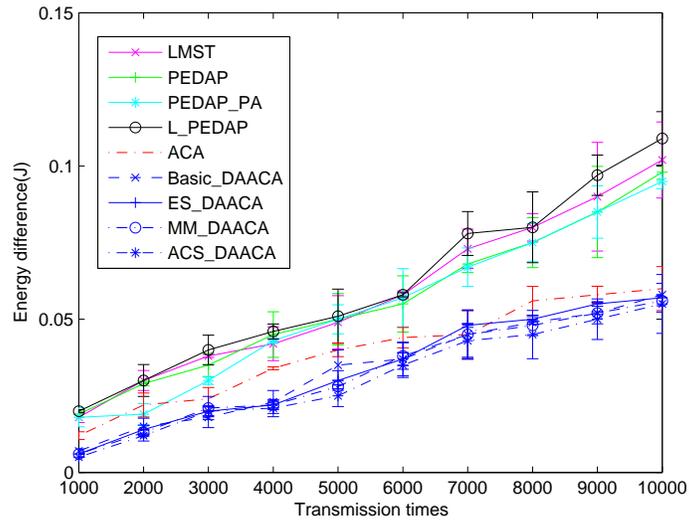}\\
  \caption{Energy Difference of the Network with the size (140 $\times$ 140, 2000)}\label{figure5}
\end{figure}

\begin{figure}
  \centering
  \includegraphics[width=10cm]{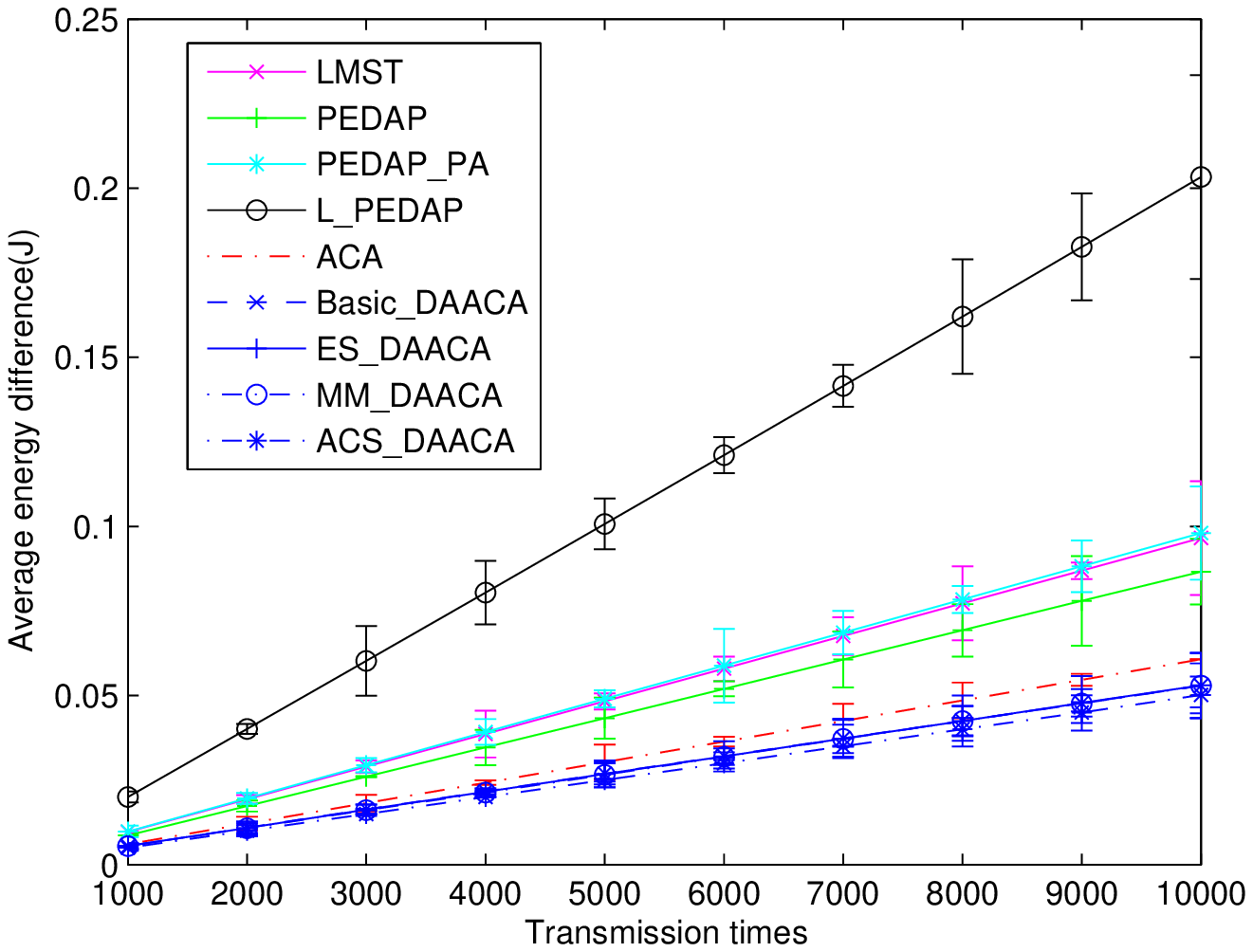}\\
  \caption{Average Results of all the Experiments of Average Energy Cost Energy Difference}\label{figure6}
\end{figure}
In figure \ref{figure4} - figure \ref{figure6}, we can see that the energy difference of L-PEDAP is the highest, the reason is that in L-PEDAP, each node needs to establish local minimum spanning tree periodically. L-PEDAP is energy aware, the weights of edges in the neighborhood always change, which demands each node to set up the local minimum spanning tree repeatedly. Although L-PEDAP can balance the consumption of energy within neighborhood, a node with bigger neighbor set will actually consumes more energy than the smaller one. Because in the dense region of the network, each node owns many neighbor nodes. When establishing the LMST, each node needs to communicate with all the neighbors to acquire the energy information of the neighbors. Therefore, a node with large neighbor node set will spend much energy in establishing LMST, whereas a node owns small neighbor set cannot spend much energy, which yields unbalanced energy consumption such two kinds of nodes. Therefore, a big energy difference occurs.

Although LMST utilize RNG to establish the network topology, the same problem of L-PEDAP still exists, which leads to a big energy difference.

In PEDAP and PEDAP-PA, the sink node consumes most of the energy, because it needs to broadcast the topology of the network to all nodes in the network periodically. Thereby, there is a large energy difference between the sink node and others.

Whereas in DAACA, it takes the advantages of both centralized and decentralized algorithms. When selecting the next hop, each node will estimate the remaining energy of its neighbor to achieve balanced usage of energy. Moreover, in the process of depositing global pheromones, the path with minimum energy cost will obtain additional pheromones, by which the balanced usage of energy is implemented globally. The results show that DAACA can balance the energy cost of the network. The energy difference relationship among DAACA family is Basic$>$MM$>$ES$>$ACS.

\subsection{Network Lifetime} \label{NetworkLifetime}
\begin{figure}
\centering
  \includegraphics[width=10cm]{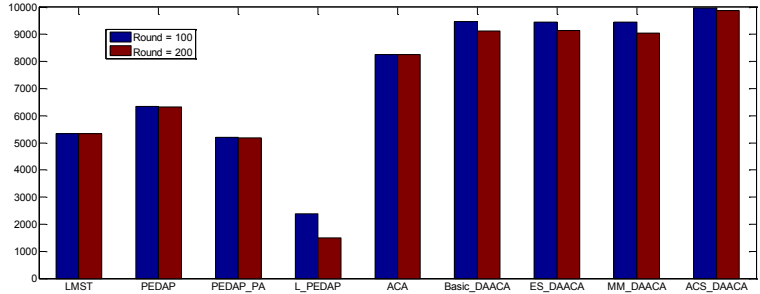}\\
  \caption{Lifetime of Network}\label{figure7}
\end{figure}

The initial energy of each node is set as 0.05J. We measure how many rounds the network will sustain until any node exhausts in each algorithm. The parameter $roundToUpdate$ is set as 100 and 200 respectively. From figure \ref{figure7}, the DAACA family shows longer lifetime than the other algorithms. And the relationship within the DAACA family is ACS$ > $MM$ > $ES$ > $Basic.
In PEDAP and PEDAP-PA, the sink node will firstly exhausted because of frequent broadcasting of topology. The lifetime of L-PEDAP and LMST are short which mainly results from massive energy costs in establishing local minimum spanning tree. In ACA, the energy cost mainly concentrates on some shortest paths. DAACA can guarantee the balanced energy costs of nodes by adjusting quantities of pheromones, therefore, it can prolong the lifetime of the network.

\subsection{Successful Ratio of One Hop Transmission}
\begin{figure}
  \centering
  \includegraphics[width=10cm]{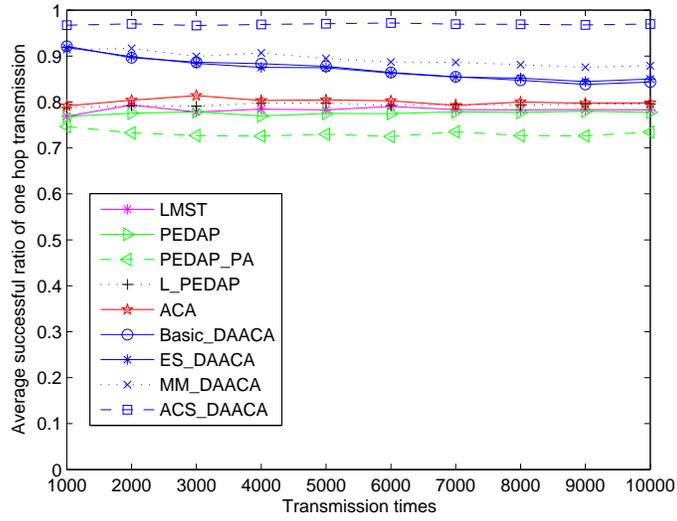}\\
  \caption{Average Successful Ratio of One Hop Transmission}\label{fig:success}
\end{figure}

In this section, we come to analyze the average success ratio of one hop transmission of each algorithm. We choose the data set ($100\times 100$, 1000). Packets are sent from the sources to the destination nodes. We count the success ratio of one hop transmission of each packet and calculate the average successful ratio of the one hop transmission. In figure \ref{fig:success}, the successful ratio of DAACA is higher than other algorithms. The relationship within the DAACA family is ACS$>$MM$>$ES$>$Basic.

\subsection{Network Topology Analysis} \label{NetworkTopologyAnalysis}
In this section, we analyze the performance of network topology in terms of node average degree and average transmission radius. As the experiment in \cite{label-7}, the data set (100 $\times$ 100, 1000) is chosen to run the simulation.

\subsubsection{Average Degree}
In this experiment, we come to analyze the average degree of each node. The average degree is calculated as the number of edges divided by the number of nodes. Bigger average degree indicates the higher spatial resue\cite{label-7}.
\begin{table}[!hbp]
\centering
\caption{Average Degree} \label{AverageDegreeTable}
\begin{tabular}{cc}
\toprule
\textbf{Algorithm} & \textbf{Average Degree} \\
\midrule
LMST & 2.04 \\
PEDAP \& PEDAP-PA & 0.999 \\
L-PEDAP & 2.04 \\
ACA & 0.999 \\
DAACA & 0.999 \\
\bottomrule
\end{tabular}

\end{table}

Table \ref{AverageDegreeTable} lists the average degree of each node in the network. In PEDAP and PEDAP-PA, the Prim algorithm is used to generate the minimum spanning tree of the network globally, since in the minimum spanning tree, the number of edge equals to the number of nodes minus 1, therefore the sum of the degree equals to the number of nodes minus 1 and the average degree is $n-1/n$ where $n$ is the number of nodes in the network. In ACA and DAACA, although the topologies are dynamic, each time only one node is selected as the next hop until the packet arrives at the sink node. Therefore the total degree still equals to that of PEDAP and PEDAP-PA. Whereas in LMST and L-PEDAP, local MST is produced, which cause many loops in the topology, in such circumstance the number of edges is bigger than $n$. Hence, the average degree is higher than other algorithms.

In a word DAACA can achieve a higher spatial reuse.

\subsubsection{Average Transmission Radius} \label{AverageTransmissionRadius}
In this experiment, we calculate the average transmission radius of each node. As illustrated in \cite{label-7}, average transmission radius is another factor which indicates the spatial reuse.

\begin{table}[!hbp]
\centering
\caption{Average Transmission Radius} \label{AverageTransmissionRadius}
\begin{tabular}{cc}
\toprule
\textbf{Algorithm} & \textbf{Radius} \\
\midrule
LMST & 4.81 \\
PEDAP \& PEDAP-PA & 2.19 \\
L-PEDAP & 2.07 \\
ACA & 1.8 \\
DAACA & 1.79 \\
\bottomrule
\end{tabular}

\end{table}
As illustrated in \cite{label-7}, the smaller average transmission radius will lead to a higher spatial reuse. In table \ref{AverageTransmissionRadius}, the results of the average transmission radius of each algorithm are listed. In LMST and L-PEDAP, the local MST are established, more links will be used to establish the topology, which is the main reason of causing the higher transmission radius. In PEDAP and L-PEDAP, the average degree is 2.19 and 2.07 respectively. In ACA and DAACA, the next hop node is selected which is nearer to the sink node than to the current node. The selection is based on the probabilities; the shorter edge has a higher possibility to be chosen. In DAACA the edge is selected based on distance and energy, therefore, the radius is a little shorter than ACA.

From the topology analysis, we can conclude that DAACA achieves a higher spatial reuse than the other algorithms.

\subsection{Time Complexity Analysis}
In this section, we mainly analyze the complexity of constructing and maintaining the topology. Because in the initialization phase, the steps (e.g. initialize the neighbor set, initialize routing protocol, et al.) are nearly identical, we only compare the time complexity in constructing and maintaining the topology of each method.

\subsubsection{Time Complexity of Topology Construction} \label{ComplexityofTopologyConstruction}
\begin{table}[!hbp]
\centering
\caption{Complexity of Topology Construction} \label{ComplexityOfTopologyConstructionTable}
\begin{tabular}{cc}
\toprule
\textbf{Algorithm} & \textbf{Time Complexity} \\
\midrule
LMST & $O(nelog(n))$ \\
PEDAP-PA \& LEDAP-PA & $O(elog(n))$ \\
L-PEDAP & $O(nelog(n))$ \\
ACA & $O(n^2)$ \\
DAACA & $O(n^2)$ \\
\bottomrule
\end{tabular}

\end{table}

Table \ref{ComplexityOfTopologyConstructionTable} reveals the time complexity of each algorithm in constructing network topology. In LMST, each node establishes the local minimum spanning tree, the Prim algorithm is utilized to generate the MST, the complexity to set up MST is $O(elog(n))$, where $e$ is the number of links within the visible neighborhood \cite{label-8} and $e \geq n$. But LMST is only apt to the network with a small number of nodes or low density network. Because when LMST is taken upon numerous nodes, each node will own a large set of neighbors which yields to $e \approx n^2$. Therefore the overall complexity in constructing LMST is high. Whereas in PEDAP and PEDAP-PA, the centralized version of minimum spanning tree is established, which owns a complexity of $O(elog(n))$. Compared with LMST, the complexity is relatively low, but it is a centralized method which casts a heavy burden on the sink node. In L-PEDAP, the topology construction is identical with LMST; therefore, the complexity is still high. In ACA, when establishing the network topology, each node must broadcast messages to his neighbors in order to inform them its hops to the sink node. Therefore, the complexity is $O(n^2)$. Topology construction DAACA is nearly the same as ACA, thus the complexity is $O(n^2)$.

\subsubsection{Time Complexity of Topology Maintenance} \label{ComplexityOfTopologyMaintenance}
In this section, the complexity of maintaining the topology is discussed. Table \ref{ComplexityOfTopologyMaintenanceTable} lists the complexity of each algorithm. Since each method updates its topology in every $roundeToUpdate$ rounds periodically, it is feasible to compare the complexity in just one period. In LMST, PEDAP, PEDAP-PA and L-PEDAP, all the topologies are required to be re-constructed. Thus the complexities are the identical with those in constructing topologies of section \ref{ComplexityofTopologyConstruction}. In ACA, whenever a node meets the condition of $EHC \leq TTL$, operations of pheromones are performed, some extra control messages \cite{label-10} are needed to be transferred, therefore, the complexity of ACA is $O(rn)$ ($r$ stands for $roundToUpdate$). Whereas in the family of DAACA, each node locally calculates the probability of selecting the next hop in every round. Only one broadcasting message for updating the energy evaluation is required, therefore, the complexity is $O(n)$. From table \ref{ComplexityOfTopologyMaintenanceTable}, it can be concluded that the complexity of maintaining the topology of DAACA is lower than the other methods.

\begin{table}[!hbp]
\centering
\caption{Complexity of Topology Maintenance} \label{ComplexityOfTopologyMaintenanceTable}
\begin{tabular}{cc}
\toprule
\textbf{Algorithm} & \textbf{Time Complexity} \\
\midrule
LMST & $O(nelog(n))$ \\
PEDAP-PA \& LEDAP-PA & $O(elog(n))$ \\
L-PEDAP & $O(nelog(n))$ \\
ACA & $O(rn)$ \\
DAACA & $O(n)$ \\
\bottomrule
\end{tabular}

\end{table}

\subsection{Other Characteristic}
In this section, we analyze other characteristics of DAACA.
\subsubsection{Robustness} \label{roubustness}
When selecting the next hop, each node will refer to the probability recorded in the routing table, all the path are dynamical, if a node is removed from the network, only the nodes nearby are required to update the routing table. Therefore, the adjustment of removing a node can be achieved locally, therefore, DAACA owns the feature of robustness.

\subsubsection{Fault Tolerance} \label{faultTolerance}
Suppose errors occur in the element of the node which prohibits it from working. In this case, the node nearby can remove this node from the routing table. Therefore, the faulty nodes will no longer exist in the topology of the network. Hence they cannot affect the transmission of the network. And we can conclude that DAACA is fault tolerant to faulty nodes.

\subsubsection{Scalability} \label{scalability}
Since nodes are deployed randomly in the network, and each node only needs to maintain the routing table for dynamically selecting the next hop. Therefore, if a node is added in the network, it only needs to broadcast his identity information and sets up routing table, the nodes nearby are only required to add an new item of the routing table. So we can conclude that DAACA is scalable.

\section{Conclusion} \label{conclusion}
In this paper, a family of ant colony algorithms called DAACA for data aggregation has been presented which contains three phases: the initialization, packet transmission and operations on pheromones. After initialization, each node estimates the remaining energy and the amount of pheromones to compute the probabilities used for dynamically selecting the next hop. After certain rounds of transmissions, the pheromones adjustment is performed periodically, which combines the advantages of both global and local pheromones adjustment for evaporating or depositing pheromones. Four different pheromones adjustment strategies are designed to achieve the global optimal network lifetime, namely Basic-DAACA, ES-DAACA, MM-DAACA and ACS-DAACA. Compared with some other data aggregation algorithms, DAACA shows higher superiority on average degree of nodes, energy efficiency, prolonging the network lifetime, computation complexity and success ratio of one hop transmission. At last we analyze the characteristic of DAACA in the aspects of robustness, fault tolerance and scalability.

\section*{ACKNOWLEDGEMENTS}
This work was partially supported by the National Natural Science Foundation of China under Grants No.60703101 and No.60903153, and the Fundamental Research Funds for the Central Universities.

\bibliographystyle{elsarticle-num}
\bibliography{DAACA}

\end{document}